\documentstyle[12pt,a4]{article}
\begin{document}
\normalsize

\title{\bf Thompson's Method applied to Quantum Electrodynamics (QED)}

\author{Cl\'audio Nassif and P.R. Silva}

\thanks{ cnassif@fisica.ufmg.br\\ \\
Departamento de F\'{\i}sica - ICEx - UFMG\\
Caixa Postal 702 - 30.123-970 - Belo Horizonte - MG - Brazil \\
}

\maketitle
\begin{abstract}
\noindent
In this work we apply Thompson's method (of the dimensions) to study the quantum electrodynamics (QED). This method can be considered as a simple and alternative way to the renormalisation group (R.G) approach and when applied to QED lagrangian is able to obtain the running coupling constant behavior $\alpha (\mu)$, namely the dependence of $\alpha$ on the energy scale. We also obtain the dependence of the mass on the energy scale. The calculations are evaluated just at $d_c=4$, where $d_c$  is the upper critical dimension of the problem, so that we obtain logarithmic behavior both for the coupling $\alpha$ and the mass $m$ on the energy scale $\mu$.\\
\\
PACS Number(s): 12.20
\end{abstract}

\section{Introduction}

\hspace{3em}

There are a considerable number of problems in Science where fluctuations are present in all scales of length, varying from microscopic to macroscopic wavelengths.

As examples, we can mention the problems of fully developed turbulent fluid flow, critical phenomena and elementary particle physics. The problem of non-classical reaction rates (diffusion limited chemical reactions) turns out also to be in this category.

As was pointed out by Wilson \cite{1}: ``in quantum field theory, ``elementary" particles like electrons, photons, protons and neutrons turn out to have composite internal structure on all sizes scales down to zero. At least this is the prediction of quantum field theory".

The most largely employed strategy for dealing with problems involving many length scales is the ``Renormalization - Group (RG) approach".
The RG has been applied to treat the critical behavior of a system undergoing second order phase transition and has been shown to be a powerful method to obtain their critical indexes \cite{2}.

Refering to the Quantum Electrodynamics (QED), Gell-Mann and Low \cite{3} obtained a RG equation for electron charge $e_\mu$, being $\mu$ the energy scale, so that in the limit as $\mu$ goes to zero we obtain the classical electron charge {\sl e}, and as $\mu$ goes to infinity we get the bare charge of the electron $e_B$.

The differential equation evaluated by Gell-Mann and Low obtains the ``experimental" charge $e_\mu$ of electron as a function of the energy, which corresponds to an interpolation between the classical and bare charges, namely:$e< e_\mu < e_B$.

In an alternative way to the RG approach, C. J. Thompson \cite{4} used a heuristic method (of the dimensions) as a means to obtaining the correlation length critical index ($\nu$), which governs the critical behavior of a system in the neighborhood of its critical point. Starting from Landau-Ginzburg-Wilson hamiltonian or free energy, he got a closed form relation for $\nu(d)$ \cite{4}, where d is the spatial dimension. It is argued that the critical behavior of this $\Phi^4$-field theory is within the same class of universality as that of the Ising Model.

Recently one of the present authors \cite{5} applied Thompson's method to the study of the diffusion limited chemical reaction ${\bf A+A \to 0}$ (inert product). The results obtained in that work \cite{5} agree with the exact results of Peliti \cite{6} who renormalized term by term given by the interaction diagramms in the perturbation theory.

More recently, Nassif and Silva \cite{7} proposed an action to describe diffusion limited chemical reactions belonging to various classes of universality. This action was treated through Thompson's approach and could encompass the cases of reactions like ${\bf A+B \to 0}$ and  ${\bf A+A \to 0}$ within the same formalism. Just at the upper critical dimensions of  ${\bf A+B \to 0}$ ($d_c=4$) and  ${\bf A+A \to 0}$ ($d_c=2$) reactions, the present authors found universal logarithmic corrections to the mean field behavior.

Thompson's method has been applied to obtain the correlation length critical exponent of the Random Field Ising Model by Aharony, Imry and Ma \cite{8} and by one of the present authors \cite{9}. His method was also used to evaluate the correlation critical exponent of the N-vector Model \cite{10}. Yang - Lee Edge Singularity Critical Exponents \cite{11} has been also studied by this method.

The aim of this work is to apply Thompson's method to the study of the QED. We intend to evaluate the QED coupling  $\alpha$  and the renormalized electron mass $m$ as functions of the energy scale $\mu$, obtaining logarithmic corrections as a function  of the energy scale $\mu$ for coupling $\alpha$ and electron mass. In order to do this we are going to define an ``Interpolated Lagrangian" $\mathcal{L}_I$ in terms of a charge parameter ${e^2_I}=\alpha_I$, being $\alpha_I$ a finite quantity so that $\alpha \le {\alpha_I} < \alpha_B$, where ${\alpha_I} \approx \alpha$ in low energy scales and ${\alpha_I} = \alpha_{min} = \alpha_0 \approx \frac {1}{137}$ as $\mu \to 0$.

We start in section II, by first considering a heuristic prescription which characterizes Thompson's method \cite{4}. We introduce the QED lagrangian, looking at it under two differents viewpoints. These are reminscent of the usual way of treating the QED, where we have the bare lagrangian ($\mathcal{L}_B$) and the renormalized lagrangian ($\mathcal{L}$). However the two lagrangians we are going to define here have finite parameters which differ through multiplicative functions depending uppon the energy scale.
In this section we also relate the parameters defining the interpolated lagrangian ${\mathcal{L}}_I$ to the physical ones. ${\mathcal{L}}_I$ reseambles the $\mathcal{L}_B$, but we keep their parameters as finite quantities.

Third section is dedicated to some further elaboration of Thompson's method when applied to QED.
 
In the fourth section we study the equation describing the dependence  both of the coupling constant $\alpha$  and the mass $m$ on the energy scale ($\mu$) where $\alpha_I$ (the interpolated coupling constant) appears as a free parameter. We consider two approximations which are based on certain plausible hypothesis relating  $\alpha_I$ to $\alpha$. In the first approximation, by putting ${\alpha_I} \approx \alpha$, we get a differential equation for $\alpha$, which agrees with the results for the renormalized perturbative expansion of QED when performed at one loop level. In the second approximation we treat the case of higher energy scales where ${\alpha_I} \gg \alpha$, and a relation between these two couplings will be proposed as a means to obtain a new differential equation for $\alpha$. The effects of these two approximations on the mass $m(\mu)$ are also discussed.

The last section is dedicated to the conclusions and prospects.
\section{QED Lagrangian under the Thompson's \\
  method viewpoint}
\hspace{3em}
We start this section by writting the physical QED lagrangian, namely: 
\begin{equation}
{\mathcal{L}} = i{\overline{\Psi}}\gamma^\mu \partial_\mu\Psi -     m{\overline{\Psi}}\Psi-\frac{1}{4}F^{\mu\nu}F_{\mu\nu}+
ie\overline{\Psi} {\gamma^\mu} {A_\mu} {\Psi} 
\label{2.0}
\end{equation}
\noindent
with

\begin{equation}
F^{\mu\nu} = \partial^\mu A^\nu - \partial^\nu A^\mu \quad and~~~~ \overline{\Psi}=\Psi^\dagger\gamma^\circ
\label{2.0}
\end{equation}

In(1) $\Psi$ are fermion fields, {\bf \sl e} and {\bf \sl m} are respectively electron charge and rest mass, $A_\mu$ is the four-vector electromagnetic potential and $\gamma^\mu$ are the Dirac's matrices.

A heuristic prescription proposed by Thompson \cite{4} states that:\\

``When we consider the integral of the lagrangian (1) in a coherence volume $l^d$ in d-dimensions, the modulus of each integrated term of it is separately of the order of unity".

This method was firstly applied by Thompson \cite{4} to the Landau-Ginzburg-Wilson free  energy or hamiltonian, obtaining critical exponents within the same universality class of the Ising model.

But in fact, when we consider the integrals of each term in (1) as of the order of unity, we are really making a certain scaling dimensional analysis in each term of it. In doing this we have performed some scaling averages obtained separately from each integrated term of the lagrangian.

We borrows Thompson's idea to apply it to QED, but in order to better do this, we make the following considerations:

As we want the terms which appear in (1) to be quadratic forms like $\overline{\Psi}\Psi$ and $F^{\mu\nu}F_{\mu\nu}$, we are going to consider Thompson's prescription in a strong form, so that we put the modulus of each integrated term exactly equal to the unity. Besides this we will perform the integrations in the four-dimensional (4-D) euclidian space-time, in order to be consistent with the special relativity theory.

The idea of dimensional analysis is very common and was applied by Ryder \cite{12} to evaluate the dimension of $\mathcal{L}$ in QED ([${\mathcal{L}}]=l^{-d}=\Lambda^d$, in d-dimensions, where $l$ is the length and $\Lambda$ is the momentum). By applying this prescription \cite{12} to each term of $\mathcal{L}$, he obtains from the first one the dimension of the field $\Psi^2$, that is $[\Psi^2]=\Lambda^{d-1}$, which gives $[\Psi^2]=\Lambda^3=l^{-3}$ for $d=4$.

In a similar way, he \cite{12} got from the third term of $\mathcal{L}$   $[A^2_\mu]=\Lambda^{d-2}$, being $[A^2_\mu]=\Lambda^2=l^{-2}$ in the case $d=4$.

So the Thompson's approach is based on a dimensional analysis in the energy scale plus some additional heuristic conditions which lead to some mean values on the scale $l$ for the field $\Psi, A_\mu$, mass and charge (coupling $\alpha$).

Now applying Thompson's assumption in its strong form to the first term of (1) we have:

\begin{equation}
\left|\int_{l^4}i[\partial_\mu [\overline{\Psi}\Psi]]d^4x\right| = 1
\label{2.0}
\end{equation}

We can observe that the dimension of \'~$\gamma^\mu\partial_\mu$\`~                 ( $[\gamma^\mu\partial_\mu]$ ) which would appear in the integral is the same as $[\partial_\mu]=l^{-1}$ because we are thinking only about a dimensional analysis in (3) for \'~$\gamma^\mu\partial_\mu$\`~. So we can neglect the spinorial aspect of the field and just consider the \'~first derivative 
$\partial_\mu$\`~ which defines the fermions regarding to the scaling dimensional analysis \cite{12}.

On the other hand, when we are dealing with scalar fields, the second derivative picks up the bosonic behavior in a scaling dimensional analysis.

We would like to stress that, in his treatment of the critical phenomena, Thompson \cite{4} has considered integrals of the kind given by (3) in a more general case of d-dimensions. However, our interest here is restricted to the four-dimensional case, which is the most relevant if we take in account relativity theory, namely QED in (3+1) dimensions.

It is interesting to note that the integral (3) leads to a kind of scaling dimensional analysis \cite{12}, where the dimensional value of certain quantity $[\overline{\Psi}\Psi]$ inside the integral is taken out of its integrand as a mean value in a coherent volume $l^4$. Then, from (3) we have:\\
$\left<[\overline{\Psi}\Psi]\right>_l \equiv [\overline{\Psi}\Psi]_l \sim l^{-3}$, which corresponds a mean value of $[\overline{\Psi}\Psi]$ on scale $l$, where we considered a 4-D cubic volume $l^4$ for (3).

In order to apply Thompson's prescription to the second term of (1), we would consider the relation $\left|-\int_{l^4}[m\overline{\Psi}\Psi]d^4x\right| = 1$. However, a close examination reveals that this procedure does not work quite well. Due to the coupling between the $\Psi$ and $A$ fields, it is the mass increment that must be considered in the above relation. By putting $\Delta m=m(\mu)-m_0$, being $\mu$ the energy scale ($\mu = l^{-1}$), $\Delta m$ goes to zero as $\mu \to 0$ (or equivalently $l \to \infty$). After these considerations we can write: 
\begin{equation}
\left|-\int_{l^4}[\Delta m][\overline{\Psi}\Psi]d^4x\right| = 1
\label{2.0}
\end{equation}

It is worth to emphasize that relation (4) was written making the requirement that the quantities involved in it must satisfy a scaling relation. Relation (4) implies that:
\begin{equation}
\left<[\Delta m]\right>_l\left<[\overline{\Psi}\Psi]\right>_ll^4=1~,
\label{2.0}
\end{equation}
\\
or
\begin{equation}
[\Delta m]_l[\overline{\Psi}\Psi]_ll^4=1
\label{2.0}
\end{equation}

As we know, $[\overline{\Psi}\Psi]_l$ goes as $l^{-3}$,such that from (5) we get that $[\Delta m]_l$ scales as $l^{-1}$. This result is consistent with
 $[\Delta m]_l \to 0$ as $l \to \infty.$
 
To use Thompson's assumption in the third term of (1), that is better to think in terms of the density of energy in the electromagnetic field ($\rho$), that is to say:\\

\begin{equation}
\frac{1}{4\pi}\int_{l^4}[|E^2|]d^4x=\frac{1}{4\pi}\int_{l^4}[|B^2|]d^4x = 1~,
\label{2.0}
\end{equation}
\\
where we have used $|E^2|=|B^2|$, being $\rho=\frac{1}{8\pi}(|E^2|+|B^2|)=\frac{1}{4\pi}|E^2|=\frac{1}{4\pi}|B^2|$.

Relation (7) implies that $[|E^2|]_l = [|B^2|]_l \sim l^{-4}$. We know that $\vec B=\nabla \times \vec A$, so making a dimensional analysis for A, we obtain: $[A^2]_l=[|B^2|]_ll^2 \sim l^{-4}l^2=l^{-2}$.

With respect of the last term of lagrangian (1) it is better to evaluate the average of it on the scale $l$ by considering a term which is quadratic in the electromagnetic potencial $A_\mu$. This choice could be justified taking in account that the interaction between electric charges is mediated by virtual photons, and the energy density of the electromagnetic field goes with $[|B^2|]$ which behaves as $[A^2]l^{-2}$ in terms of dimensional analysis. Besides this it must be stressed that our central interest here is in the evaluation of coupling constant $\alpha = e^2$,instead of the pure electric charge $e$. So we must think an effective contribution for the action through  a product of integrals, namely we must look for a product of integrals in a 4-dimensional space, which corresponds to an average of the square of the last term of (1) in a space of 8-dimensions. Based on these considerations we write:

\begin{equation}
\left|i^2\int_{l^{\prime 4}}(\int_{l^4}[e[(\overline{\Psi}\Psi)]A_\mu][e^{\prime}[(\overline{\Psi^{\prime}}\Psi^{\prime})]A^{\prime}_\mu] d^4x )d^4x^{\prime} \right| = 1~,
\label{2.0}
\end{equation}
\\
where \'~{$\prime$}~\`~ is a dummy index and we can take: $d^4x^{\prime}d^4x=d^8x$.

So we can write (8) in the following way:

\begin{equation}
\left|i^2\int_{l^8}\left[e^2[(\overline{\Psi}\Psi)]^2A^2_\mu \right] d^8x \right| = 1
\label{2.0}
\end{equation}

As one step further in the direction of obtaining the differential equation describing the running coupling constant of the QED, let us define the inner lagrangian $\mathcal{L}_I$, whose parameters ($\alpha_I, m_I, \Psi_I$ and $A^I_\mu$) interpolate the parameters of the lagrangian $\mathcal{L}$ and the bare lagrangian $\mathcal{L}_B$, such that $\alpha(\mu) \le \alpha_I<\alpha_B$. $\mathcal{L}_I$ is constructed so that at lower energy scales we obtain the following approximations: $\alpha_I \approx \alpha$ and $m_I \approx m$. So we write:

\begin{equation}
{\mathcal{L}}_I = i{\overline{\Psi_I}}\gamma^\mu \partial_\mu\Psi_I -     m_I{\overline{\Psi_I}}\Psi_I-\frac{1}{4}F^{\mu\nu}_IF^I_{\mu\nu}+
ie_I\overline{\Psi_I} {\gamma^\mu} {A^I_\mu} {\Psi_I}~. 
\label{2.0}
\end{equation}

We keep the parameters defining ${\mathcal{L}}_I$ as finite quantities, although being functions which increase with increasing energy. Vacuum polarization shields the inner charge $e_I$ in such a way that the measured (physical) coupling constant ($\alpha$) is always smaller than the inner parameter $\alpha_I$. We suppose that at higher energy regime the inner charge is substantially greater than the measured charge due to the increasing of vacuum polarization while at lower energies they are approximately equals because the vacuum polarization decreases.

We could think in terms of the scaling functions of the energy $\mu$ mapping the quantities of $\mathcal{L}$ into $\mathcal{L}_I$. However, for calculational proposes, it would be better to redefine $\mathcal{L}_I$ in the following way:

\begin{equation}
{\mathcal{L}}_I = iw_2(\mu){\overline{\Psi}}\gamma^\mu \partial_\mu\Psi -     w_0(\mu) m{\overline{\Psi}}\Psi- w_3(\mu)\frac{1}{4}F^{\mu\nu}F_{\mu\nu}+
iw_1(\mu)e\overline{\Psi} {\gamma^\mu} {A_\mu} {\Psi}~. 
\label{2.0}
\end{equation}

Identifying the lagrangians (10) and (11) by imposing the equality of them term by term, we obtain:

\begin{equation} \label{eqarray}
  \begin{array}{ccc}
  \Psi_I &=&(w_2)^{\frac{1}{2}}\Psi~~;\\
  F^{\mu\nu}_I &=&(w_3)^{\frac{1}{2}} F^{\mu\nu}~~;
  \end {array} 
  \quad
  \begin{array}{ccc}
   m_I&=&\frac{w_0}{w_2}m \\
  e_I &=& \frac{w_1}{w_2w^{\frac{1}{2}}_3} e
  \end{array}
\end{equation}

We must emphazise that $w_0$, $w_1$, $w_2$ and $w_3$ are functions which increase monotonically with the energy $\mu$. Therefore they will go to infinity only when $\mu \to \infty$.

\section{Some further elaboration of Thompson's method applied to QED.}

\hspace{3em}
Let us now to consider integral (3) evaluated in a volume of a 4 - D hyper-sphere, once we are interested in the isotropic 4 - D space-time, being the scale of length $l$ the radius of this hiper-sphere.

The volume of a n - D hyper-sphere is given by $V_n = S_n\frac{l^n}{n}$ \cite{13} where $S_n =\frac{2\pi^{\frac{n}{2}}}{\Gamma(\frac{n}{2})}$ \cite{13}. In 4 - D, we have $V_4 = \frac{\pi^2l^4}{2}$, implying in $dV_4 = 2\pi^2 r^3 dr$, with $r$ a radial variable.

The above considerations permit us to write integral (3) as:
  
\begin{equation}
   2\pi^2\int_{0_{(V_4)}}^l \partial_r [\overline{\Psi}\Psi]_r r^3dr =       \left<[\overline{\Psi}\Psi ]\right>_l 6\pi^2\int_{0}^l r^2dr=1
\label{2.0}
\end{equation}

Equation (13) implies that:\\
  \begin{equation}
    \left<[\overline{\Psi}\Psi] \right>_l \equiv [\overline{\Psi}\Psi]_l = \frac{1}{2\pi^2l^3}~,
  \label{2.0}
   \end{equation}
where $2\pi^2l^3$ is the magnitude of the surface of this 4-D hyper-sphere.

$[\overline{\Psi}\Psi]_l$ could be thought of as a mean condensate of fermions where the average is taken on a length scale $l$, being $l=\mu^{-1}$. Therefore this condensate has the dimension of $\mu^3$ (the third power of energy). The $2\pi^2$ constant is a consequence of the spherical simmetry we have assumed for the problem.

By putting the first transform defined in (12), namely $\Psi=\Psi_I (w_2)^{-\frac{1}{2}}$ in (14) we obtain

  \begin{equation}
    \left<[\overline{\Psi_I}\Psi_I] \right>_l \sim \frac{\left<w_2 \right>_l}{2\pi^2l^3}~,
  \label{2.0}
  \end{equation}
where $\left<w_2\right>_l$ corresponds to an average of $w_2$ on the scale of length $l$. We observe that ${\left<w_2\right>}_l$ is an increasing function of the energy $\mu$ and at low energy scales it approaches to one, so that (15) recovers (14) in this regime.

Now let us  evaluate  the mass term given by integral (4) in the volume of 4-D hyper-sphere of radius $l$. We  have:

 \begin{equation}
 \left|-2\pi^2\int_{0}^l[\Delta m]_r[\overline{\Psi}\Psi]_r r^3dr\right| = 1~.
 \label{2.0}
 \end{equation}

Relation (16) implies that

 \begin{equation}
 \left<[\Delta m]\right>_l \left<[\overline{\Psi}\Psi]\right>_l \frac{\pi^2 l^4}{2} = 1~.
 \label{2.0}
 \end{equation}

By putting (14) into (17), we get:

 \begin{equation}
 \left<[\Delta m]\right>_l \equiv [\Delta m]_l = 4l^{-1}
 \label{2.0}
 \end{equation}

Relations (12), (15) and (16) imply that:

 \begin{equation}
 [\Delta m_I]_l \sim \frac{\left<w_0\right>_l}{\left<w_2\right>_l} [\Delta m]_l
 \label{2.0}
 \end{equation}

In the low energy regime, both $\left<w_0\right>_l$ and $\left<w_2\right>_l$ go to one, so that $[\Delta m_I]_l$ approaches to $[\Delta m]_l$.

Now let us consider the third term of lagrangian (1) and by performing the integrals (7) in the volume of a 4-D hyper-sphere of radius $l$ ($dV_4=2\pi^2r^3dr$), we have:

 \begin{equation}
  \frac{\pi}{2}\int_{0}^l[|E^2|]_r r^3dr = \frac{\pi}{2}\int_{0}^l[|B^2|]_r        r^3dr =1
 \label{2.0}
 \end{equation}

Relations (20) leads to

  \begin{equation}
  [|E^2|]_l  = [|B^2|]_l = \frac{8}{\pi l^4}
 \label{2.0}
 \end{equation}

From the definition  $\vec B=\vec \nabla \times \vec A$, we are led to the scaling relation for $[A^2]_l$:

 \begin{equation}
  [B^2]_l l^2  = [A^2]_l = \frac{8}{\pi l^2}~~,
 \label{2.0}
 \end{equation}

where in obtaining (22), we have used (21).

It is interesting to note that (22) is consistent with a potential of a static point charge, that is to say $\Phi \sim \frac{1}{r}$, which leads to $[\Phi^2]_l \sim \frac{1}{l^2}$, where $\Phi \equiv A_4$ and $A_{\mu} = (\vec A, \Phi)$. These considerations permit us to write (22) in a compact form:
 
 \begin{equation}
  [A^2_{\mu}]_l   =  \frac{8}{\pi l^2}~~,
 \label{2.0}
 \end{equation}

With respect to the fourth term of $\mathcal{L}$ in (1), previous considerations had led to integral given by (9). Therefore let us evaluate integral (9) in a volume of a 8-D hyper-sphere. Taking in account that $d^8x \equiv dV_8 = S_8r^7dr=\frac{\pi^4}{3}r^7dr$, we have:
  
 \begin{equation}
 \frac{\pi^4}{3}\int_{0}^l[\alpha]_r[\overline{\Psi}\Psi]^2_r [A^2_{\mu}]_rr^7dr = 1~,
 \label{2.0}
 \end{equation}
\noindent
where $\alpha = e^2$.

A first trying in order to evaluate (24) could be to write it as a product of averages, namely as a product of the quantities $\left<[\alpha]\right>_l, \left<[\overline{\Psi}\Psi]^2\right>_l,  \left<[A^2_{\mu}]\right>_l$ and $[V_8]_l$. By considering that $\left<[\overline{\Psi}\Psi]^2\right>_l \sim l^{-6}$, $\left<[A^2_{\mu}]\right>_l \sim l^{-2}$ and $[V_8]_l \sim l^8$, we obtain that $[\alpha]_l l^{-6} l^{-2} l^8 \sim 1$, which implies that $[\alpha]_l$ is a constant, that is to say a quantity which does not exhibit a dependence on the scale of length $l$ (or energy $\mu$): $[\alpha]_l \sim l^0 \sim$ constant.

However we must consider that $d=4$ corresponds to a kind of upper critical dimension for QED. In other words, below $d=4$  fluctuations are very important to the problem, and above $d=4$, ``mean field" description is a good description to the problem. So $d=4$ represents a border-line dimension for QED and we must improve our approximations in order to ``see" the dependence of the coupling $[\alpha]_l$ on the length scale $l$, or equivalently on the energy scale $\mu=l^{-1}$. A similar situation has been occurred when we treated diffusion limited chemical reactions through Thompson's method \cite{5,7}. As a means to improve the calculation of (24) let us take the quantities $[\overline{\Psi}\Psi]^2_r$ and $[A^2_{\mu}]_r$ inside the integral with the same form as that evaluated in (14) and (23), but displaying a dependence on the r-variable of scale. By taking inside the integral (24) $[\overline{\Psi}\Psi]_r = \frac{1}{2\pi^2 r^3}$ and $[A^2_{\mu}]_r = \frac{8}{\pi r^2}$, we can write : 
  
\begin{equation}
 \frac{\pi^4}{3}\int_{l}[\alpha]_r \left(\frac{1}{2\pi^2r^3}\right)^2 \left(   \frac{8}{\pi r^2}\right) r^7dr = 1
 \label{2.0}
\end{equation}

From (25), we have:

\begin{equation}
 [\alpha]_l\frac{2}{3\pi}\int_{1}^l \frac{dr}{r} = [\alpha]_l \frac{2}{3\pi}ln(l) = 1
 \label{2.0}
\end{equation}

In evaluating (26), we have taken  1 as a lower cutoff on the scale $l$. Therefore (26) displays the logarithmic dependence for $[\alpha]_l$ on the scale of length $l$ (or energy $\mu = l^{-1}$).

However it is possible to go deeper  into this subject by considering the inner lagrangian $\mathcal{L}_I$ and the transforms given by (12). Since we have from (12) $ \alpha_I = \frac{w^2_1}{w^2_2w_3} \alpha$, by inserting this relation into  (25) we obtain 

\begin{equation}
 [\alpha_I]_l\frac{2}{3\pi}\int_{l} \frac{1}{r}dr = \left< \frac{w^2_1}{w^2_2w_3}\right>_l~~.
 \label{2.0}
\end{equation}

It is convenient to write (27) in the form:

\begin{equation}
 [\alpha_I]_l\frac{2}{3\pi}\int_{l} \frac{1}{r}dr = \left< \frac{[\alpha_I]}{[\alpha]}\right>_l~~.
 \label{2.0}
\end{equation}

Equation (28) is more general and contains in principle the behavior of the couplings on all the energy scales, obtained in a non-perturbative closed form. At low energies $\alpha_I \to \alpha$ and $\left< \frac{[\alpha_I]}{[\alpha]}\right>_l \to 1$, and in this way from (28) we recover (26). For higher energy scales, we need to conceive a function representing the rapid increasing of $[\alpha_I]_l$ with respect the physical measured value $[\alpha]_l$, where at higher energies we are supposing $[\alpha_I]_l \gg [\alpha]_l$.

\section{Evaluation of the dependence of charge and mass of the electron with the scale of energy}

 \subsection{Obtaining $\alpha(\mu)$}
\hspace{3em}
At lower energy scales in the quantum regime (vacuum polarization), the behavior of $\alpha$ is given by Eq.(26).\\
For sake of simplicity in the notation we write:

\begin{equation}
 \left<[\alpha]\right>_l \equiv [\alpha]_l \equiv \alpha(l)~~,
 \label{2.0}
\end{equation}

and by putting $\mu=l^{-1}$ into (26), we get 

\begin{equation}
 -\frac{2}{3\pi}ln(\mu)=\alpha^{-1}(\mu)~.
 \label{2.0}
\end{equation}

Differentiating both sides of (30) with respect the $\mu$ variable, we obtain:

\begin{equation}
 \mu \frac{d\alpha}{d\mu}=\frac{2}{3\pi}\alpha^2~.
 \label{2.0}
\end{equation}

Equation (31) coincides with that which is obtained by the R.G procedure, when the QED is treated through the perturbation theory at one loop level. We observe that we obtained the coefficient $\beta=\frac{2}{3\pi}\alpha^2$ \cite{14}, \cite{15}. It is worth to stress that the results of this work were obtained by using heuristic arguments, namely a dimensional analysis on the scale of length inspired in Thompson's idea \cite{4}, in a simple and alternative way to the R.G procedure.

Performing the integration of (31), by considering the limits $\mu_0$ and $\mu$ for the energy scales and their respectives couplings $\alpha(\mu_0)$ and $\alpha(\mu)$, we obtain:

\begin{equation}
 \alpha(\mu) = \frac{\alpha(\mu_0)}{1 - \frac{2}{3\pi}\alpha(\mu_0)ln\left(\frac{\mu}{\mu_0}\right)}
 \label{2.0}
\end{equation}

We observe that (32) diplays the so-called Landau's singularity, namely a finite value of the energy scale $\mu_L$, where $\alpha(\mu_L) \to \infty$.

As it is well known, Landau's singularity is a non-physical effect and reveals the fact that the running coupling constant solution given by (32) is not appropriate when the energy scale approaches to $\mu_L$.

We are led to think that at higher energies, equation (31) and its solution (32) must be modified in order to be free of the  Landau's singularity. In the usual perturbative scheme of calculation this is accomplished by considering the theory beyond one loop level (two or more loops).

Now let us look at (32). We observe that $\lim_{\mu \to 0} \alpha (\mu)=0$. But this result seems to be purely of academic interest.

Indeed even at low energy scales, the departure of the classical behavior for $\alpha(\mu)$ starts when $\mu \geq m_0$, where $m_0$ is the electron rest mass. This corresponds to assume that the effect of vacuum polarization in the shielding of electron charge becomes important when we approache to the electron closest than its Compton wavelength $l_0 \sim \lambda_c = m^{-1}_0$. Therefore, from an experimental point of view, we must look for (32) at the low energy scale regime, but with $\mu\geq\mu_0 \sim m_0$, that is, we consider the parameter of energy scale $\mu$ fixed on the electron mass $m_0$ as a scale of reference, where $\alpha(\mu_0)\sim \alpha(m_0)\approx \frac{1}{137}$. So for moderates energies we can make the expansion of (32), obtaining.

\begin{equation}
 \alpha(\mu) = \alpha_0\left[1 + \frac{2}{3\pi}\alpha_0 ln\left(\frac{\mu}{\mu_0}\right) \right]~~,
 \label{2.0}
\end{equation}
\noindent 
where  $\alpha_0 = \alpha(\mu_0) \sim \alpha(m_0)\approx \frac{1}{137}$, and $\mu_0 \sim m_0$.

As it was discussed before, for very high energies, that is to say $\mu \gg \mu_0(\sim m_0)$, approximations as (32) and (33) do not work. Therefore we need to turn to Eq.(27) as a means of trying to find the behavior of $\left< \frac{w^2_1}{w^2_2 w_3}\right>_l$ in this high energy regime. Alternatively it is better to look at the relation $\left< \frac{[\alpha_I]}{[\alpha]}\right>_l$ (see (28)). A possible way of evaluating the quantity $\left< \frac{[\alpha_I]}{[\alpha]}\right>_l$ could be to write it basically as a product of $\left<[\alpha_I]\right>_l$ times a function of $\alpha$.
We think that this quantity will be influenced at all  orders in $\alpha$, since we want to seek for the effect of high energies when the vacuum polarization increases in such a way that we have $[\alpha_I]_l \gg [\alpha]_l$. So we can guess that a good representation for it could be the following ansatz: 

\begin{eqnarray}
 \left< \frac{[\alpha_I]}{[\alpha]}\right>_l&=&\alpha^{-2}_0 \left<[\alpha_I]\right>_l e^{\left<[\alpha]\right>_l}=
\nonumber\\
     &  = &  \alpha^{-2}_0 \left<[\alpha_I]\right>_l \left[1+ \left<[\alpha]\right>_l+ \frac{1}{2!} \left<[\alpha]\right>^2_l + ....\right]~~,
 \label{2.0}
\end{eqnarray}
\noindent
where $\alpha^{-2}_0 \approx (137)^2$ would be an appropriate constant for the function behavior at higher energies scales. The exponential function above is a good representation for the rapid increasing of vacuum polarization in higher energies so that the charge shielding leads to $\left[\alpha_I\right]_l \gg \left[\alpha\right]_l$.

Inserting (34) into (28) and by considering that $\left<[\alpha]\right>_l \equiv \alpha(\mu)=\alpha$, that is $\alpha$ measured on scale, where $\mu=l^{-1}$, we get

\begin{equation}
 \frac{2}{3\pi} \alpha^2_0\int_{\mu} \frac{1}{r}dr = e^{\alpha_{(\mu)}}~~.
 \label{2.0}
\end{equation}

Performing integral (35), by considering the integration on energy scale $\mu$ since both sides of (35) must increase  monotonicaly with $\mu$, so we obtain from (35) the following equation:
\begin{equation}
 \frac{2}{3\pi} \alpha^2_0 ln(\mu)=e^{\alpha(\mu)}~~,
 \label{2.0}
\end{equation}
\noindent
where we considered a lower cutoff $\mu=1$ for integral (35).

Differentiating (36) with respect to $\mu$ leads to
\begin{equation}
 \frac{2}{3\pi} \alpha^2_0 \frac{d\mu}{\mu}=e^{\alpha}d\alpha~~.
 \label{2.0}
\end{equation}

From (37) we obtain:
\begin{equation}
 \mu\frac{d\alpha}{d\mu} = \frac{2}{3\pi} \alpha^2_0 e^{-\alpha}
 \label{2.0}
\end{equation}

Equation (38) is the differential equation for the running coupling ``constant" $\alpha$ in a higher energy regime rather than Eq.(31) which describes the behavior of it at the regime of lower energies.

It is relevant to make the integration of (37) by considering  the limits $\mu_0$ and $\mu$ on the energy scale with the respectives running coupling constants given by $\alpha_0$ and $\alpha$.

Here $\mu_0$ is the lower energy cutoff setted up by the electron rest mass ($\mu_0 \sim m_0 = \lambda^{-1}_c$) while $\mu$ is a variable energy scale such that $\mu\gg \mu_0$ and $\alpha$ is substantially greater than $\alpha_0$. Therefore by integration of (37) we obtain
\begin{equation}
 \alpha(\mu) = \alpha_0 + ln\left[1 + \frac{2}{3\pi}e^{-\alpha_0}\alpha^2_0ln\left(\frac{\mu}{\mu_0}\right)\right]~~,
 \label{2.0}
\end{equation}
\noindent
Once $e^{-\alpha_0}=e^{-\frac{1}{137}}\stackrel{\sim}{=} 1$, we can also approximate (39) by
\begin{equation}
 \alpha(\mu) \stackrel{\sim}{=} \alpha_0 + ln\left[1 + \frac{2}{3\pi}\alpha^2_0ln\left(\frac{\mu}{\mu_0}\right)\right]~~,
 \label{2.0}
\end{equation}

We know that $ln(1+x) \approx x $ when $x \ll 1$. So applying this approximation to (40) by considering lower energies, we recover exactly the solution (33) for moderates energies ($\mu \stackrel{>}{\sim} m_0$) as a particular case of (40).

\subsection{Obtainning $m(\mu)$}
\hspace{3em}
As a means to evaluate $\Delta m(\mu)$ let us  compare (4) and (8) by considering the shift $\Delta e^2=\Delta \alpha$ in (8) because $\Delta \alpha (\mu)$ must be directly proportional to the mass shift, that is, $\Delta m \propto \Delta \alpha$ so that in the very lower energies limit $\Delta m \propto \Delta \alpha \to 0$. Thus in doing that we have

\begin{equation}
 \left|\int_{V_4}(-\Delta m)[\overline{\Psi}\Psi]_xd^4x\right|=   \left|i^2\int_{V_4} \left(\int_{V^\prime_4}(\Delta e^{\prime 2})[\overline{\Psi^\prime}\Psi^\prime]_{x^\prime}[A^2_\mu]_{x^\prime}d^4x^\prime\right) [\overline{\Psi}\Psi]_x d^4x\right| = 1~~,
\label{2.0}
\end{equation}
\noindent
where we consider the shift $\Delta m = m-m_0$ and $\Delta \alpha=\alpha - \alpha_0$, being $m_0$ the electron rest mass and $\alpha_0 \approx \frac{1}{137}$ measured on energy scale of electron rest mass $\mu_0 \sim m_0$.  $m=m(\mu)$ and $\alpha = \alpha(\mu)$ (the running coupling ``constant") are given on energy scale $\mu$.

Relation (41) implies that
\begin{equation}
 \Delta m = \left|i^2\int_{V_4}(\Delta e^2)[\overline{\Psi}\Psi]_x[A^2_\mu]_xd^4x\right|~~,
\label{2.0}
\end{equation}
\noindent
where the index  $\prime$ is dummy.

By putting $d^4 x \equiv dV_4 = 2\pi^2 r^3 dr$, $[\overline{\Psi}\Psi]_x \equiv [\overline{\Psi}\Psi]_r = \frac{1}{2\pi^2 r^3}$ and $[A^2_\mu]_x \equiv [A^2_\mu]_r=\frac{8}{\pi r^2}$ [see (14) and (23)] into (42), we get
\begin{equation}
 \Delta m = \frac{8}{\pi}\left< [\Delta \alpha] \right>_l \left|\int_{l}\frac{1}{r^2}dr\right|
\label{2.0}
\end{equation}

Now let us take the notation $\left< [\Delta e^2] \right>_l = \left< [\Delta \alpha] \right>_l \equiv \Delta \alpha$.  This represents a certain mean charge shift measured on the scale $l \sim \mu^{-1}$.

By performing the integration indicated in (43) between the limits $l=\infty$  ($\mu=0$) and $\alpha_0 l_0 \sim \alpha_0 \lambda_c \sim \alpha_0 10^{-12}m \approx 10^{-14}m $ which is equivalent to the vacuum polarization regime, so in doing this, from (43) we obtain:
\begin{equation}
 \Delta m = \frac{8}{\pi}\Delta \alpha\frac{1}{\alpha_0l_0}=\frac{8}{\pi}\frac{\Delta \alpha}{\alpha_0}m_0~~,
\label{2.0}
\end{equation}
\noindent
or
\begin{equation}
 \frac{\Delta m}{m_0}=\frac{8}{\pi}\frac{\Delta \alpha}{\alpha_0}~~,
\label{2.0}
\end{equation}
\noindent
where $m_0=l^{-1}_0 \sim \lambda^{-1}_c$. Indeed we have the proportionality $\Delta m \propto \Delta \alpha$ obtained from (44), which leads to
\begin{equation}
 m=m_0+\frac{8}{\pi}\frac{\Delta \alpha}{\alpha_0} m_0~.
\label{2.0}
\end{equation}

Finally by substituting  $\Delta \alpha(\mu)$ obtained from (33) and (40) into (46) we get
\begin{equation}
 m=m_0 \left[1+\frac{16}{3\pi^2}\alpha_0 ln\left( \frac{\mu}{\mu_0} \right)     \right]
\label{2.0}
\end{equation}
\noindent 
and
\begin{equation}
 m \stackrel{\sim}{=} m_0 \left\{1+\frac{8}{\pi}\alpha^{-1}_0 ln\left[1+ \frac{2}{3\pi}\alpha^2_0 ln \left( \frac{\mu}{\mu_o}\right) \right] \right\}
\label{2.0}
\end{equation}
\noindent 
at a higher energy regime, where we can have experimentally $\alpha \sim \frac{1}{128}$, which corresponds to $\mu \sim 10^2Gev$ \cite{16}.

Relation (48) is more appropriate than (47) when we have $\mu \gg \mu_0$ ($\sim m_0$) so that it recovers (47) by considering that approximation where we have lower energies  ($\mu \stackrel{>}{\sim} \mu_0$), that is, $ln(1+x) \approx x$ for $x \ll 1$, making $x \equiv \frac{2}{3\pi}\alpha^2_0 ln \left( \frac{\mu}{\mu_0} \right)$ in relation (48). Therefore (47) is a particular case of (48) when we only consider $\mu  \stackrel{>}{\sim} \mu_0$.

We notice that the above relations are comparable to some results for $m(\mu)$ of the literature as quoted by Nottale \cite{17}, Weinberg \cite{18} and  Weisskopf \cite{19}.

\section{Conclusions and prospects}
\hspace{3em}
In this paper, Thompson's method which could be considered a simple alternative way to the RG calculations  was applied to study  QED. This was done by treating each term of the QED lagrangian in equal footing, in a dimensional analysis on the scale of length (or equivalently on the momentum-energy scale). If we analyse the scaling behavior of certain objects such that the mean condensate of fermions ($[\overline{\Psi}\Psi]_l$), the dimension of the squared vector potencial $[A^2_\mu]_l$, the ``excess" of mass $[\Delta m]_l$, and the ``excess" of charge $[\Delta \alpha]_l$,  with all these quantities evaluated at the scale of length $l$, we observe that it is possible to organize these objects within a hierarchical structure, thinking in terms of topological grounds. In this way, the mean condensate of fermions $\left< [\overline{\Psi}\Psi] \right>_l = (2\pi^2 l^3)^{-1}$ decreases as a ``surface" 3-D of a hypersphere 4-D of radius $l$, being this ``surface" immersed in the 4-D space-time.

The next object in this hierarchy corresponds to the dimension of the squared vector potencial. It is given by $[A^2_\mu]_l = 8(\pi l^2)^{-1}$, exhibiting a inverse square law on the scale of length $l$. This represents a 2-D structure also immersed in the 4-D space-time. We could think that for this object the degree of freedon have reduced by a unity. The ``excess" of mass $[\Delta m]_l=4l^{-1}$ can be thougth of as a 1-D structure immersed again in a 4-D space-time.

Finally the ``excess" of charge (coupling) $[\Delta \alpha]_l$ behaves in the zero-th order as scaling independent, namely $\alpha$ goes as $l^0$ at zero order in the calculations, and it can be considered as a 0-D structure immersed in a 4-D space-time. In short we have the ``spreading" of the condensate of fermions in a volume (3-D), the squared vector potential in a surface (2-D), the mass in a line (1-D) and the charge in a point (0-D), relating these objects of the QED to a hierarchical ordering in the topology of a 4-D space-time.

However when we  improve our calculations, the charge (running coupling `constant') pass to exhibit a logarithmic dependence of the scale of length. This could be considered as an intermediate regime between a point ($l^0 \sim$ constant) and a line ($l^1$). We interpret this as the charge accquiring a fractal character in this topological structure of the space-time, due to the influence of the quantum fluctuations introduced by  the vacuum polarization, in such a way that we have $\alpha(l) \sim [ln(l)]^{-1}=[l^0 ln(l)]^{-1}$.
These quantum fluctuations will also ``modulate" the behavior of the ``excess" of mass, namely $\Delta m(l) \sim [l^1 ln(l)]^{-1}$.

The fractal character of a quantum path was considered by Nottale \cite{20} on analysing the QED. He showed that due to the vacuum polarization the self-energy diagranms of the QED display a fractal character \cite{20}.

One merit of Thompson's approach is that it displays the scaling behavior of the physical magnitudes of the problem, and as a consequence, it allows us for instance to pick up the fractal structure of $\alpha (\mu)$ and $m(\mu)$.

Another advantage of the present method is that it can be extended to study, for example, the scalar field theories $ \Phi^3$, $\Phi^4$ and $\Phi^n$ for dimension $d$, in such a way that we can obtain a certain critical dimension $d_c$ for a given value of $n$ in $\Phi^n$-theory in a closed form. Here $d_c$ must be interpreted as a dimension where the $\Phi^n$-theory becomes renormalizable displaying logarithmic dependence of the coupling constant on the energy scale. This line of reasoning will be the subject of a forthcoming paper.

Finally one of the possibilities of the Thompson's method is to use it as a means to evaluate the condensate of quarks of the quantum chromodynamics (QCD). This matter will be treated elsewhere.  
\\
\\
{\noindent\bf  ACKNOWLEDGEMENTS ---} 
We are grateful to Prof. J. A. Helayell Neto for enlightening discussions and for revising the manuscript, to Dr. Marcos Sampaio for stimulating discussions at the previous stages of this work, and to Prof. O. Piguet for interesting discussions and ``insights" about the heuristic method of Thompson.
\\


\begin{thebibliography}{00}

\bibitem{1}   K. G. Wilson, Rev. Mod. Phys. {\bf 55}, 583 [1983].

\bibitem{2}   K. G. Wilson, Physica {\bf 73}, 119 [1974].

\bibitem{3}   M. Gell-Manm and F. E. Low, Phys. Rev. {\bf 95}, 1300 [1954].

\bibitem{4}   C. J. Thompson, J. Phys. {\bf A9}, L25 [1976].

\bibitem{5}   P. R. Silva, Phys. Stat. Sol. {\bf B179}, K5 [1993].

\bibitem{6}   L. Peliti, J. Phys. {\bf A19}, L365 [1996].

\bibitem{7}   C. Nassif and P. R. Silva, Mod. Phys. Lett. {\bf B13}, 829 [1999].

\bibitem{8}   A. Aharony, Y. Imry and S.-K. Ma., Phys. Rev. Lett. {\bf 37}, 1364   [1976]

\bibitem{9}  P. R. Silva, Phys. Stat. sol. {\bf B165}, K79 [1991]

\bibitem{10}   P. R. Silva, Phys. Stat. sol. {\bf B174}, 497 [1992]

\bibitem{11}   P. R. Silva, Phys. Stat. sol. {\bf B179}, K99 [1993]

\bibitem{12}  L. H. Ryder, Quantum Field Theory, Ch.9, Cambridge University Press. [1985]

\bibitem{13}  H. S. M. Coxeter, in: Regular Polytopes p125, Dover Publ. Inc.6, N. York [1973].

\bibitem{14}  A. Brizola, M. Sc. Dissertation, Universidade Federal de Minas Gerais, Belo Horizonte, Brazil. [1998]

\bibitem{15}  A. Brizola, O. Battistel, Marcos Sampaio and M. C. Nemes, Mod.Phys. Lett. {\bf A14} p.1509-1517. [1999]

\bibitem{16}  Laurent Nottale, in: Fractal Space-time and Microphysics, Ch.6, p.201, World Scientific Publishing Co. Pte. Ltd, Singapore, New Jersey, London, Hong Kong. [1993] 

\bibitem{17}  Laurent Nottale, in: Fractal Space-time and Microphysics, Ch.6, p.203, World Scientific Publishing Co. Pte. Ltd, Singapore, New Jersey, London, Hong Kong. [1993] 

\bibitem{18} S. Weinberg, The Quantum Theory of Fields, Vol1, p.496, Cambridge University Press, (USA) [1996].

\bibitem{19} V. F. Weisskopf, Phys. Rev. {\bf 56}, 72 [1939].

\bibitem{20}  Laurent Nottale, in: Fractal Space-time and Microphysics, Ch.5, World Scientific Publishing Co. Pte. Ltd, Singapore, New Jersey, London, Hong Kong. [1993]    

\end{thebibliography}
\end{document}